\documentclass[
    ,final            
  ]
  {aipproc}
\usepackage{bm, amsmath, amssymb, fdag}
\layoutstyle{6x9}
\newcommand{\tr}{{\rm tr}} 

\begin{document}

\title{The Mueller-Tang jet impact factor at NLO from the high energy effective action}


\author{Martin Hentschinski}{
  address={Instituto de Ciencias Nucleares, Universidad Nacional Aut\'onoma de M\'exico, Apartado Postal 70-543, M\'exico Distrito Federal 04510, M\'exico}
}

\author{Beatrice Murdaca}{
  address={Dipartimento di Fisica,   Universit\`a della Calabria, and
Istituto Nazionale di Fisica Nucleare, Gruppo collegato di Cosenza, I-87036 Arcavacata di Rende, Cosenza, Italy}
}

\begin{abstract}
  We report on recent progress in the evaluation of next-to-leading
  order observables using Lipatov's QCD high energy effective
  action. In this contribution we focus on the determination of the
  real part of the next-to-leading order corrections to the
  Mueller-Tang impact factor which is the only missing element for a
  complete NLO BFKL description of quark induced dijet events with a
  rapidity gap.
\end{abstract}

\maketitle


\section{Introduction}

Due to its large center of mass energy the LHC provides an ideal
opportunity to test BFKL-driven observables \cite{Fadin:1975cb, bal}.
Among them both central production processes, such as heavy quark
production \cite{heavy}, forward production of high $p_T$ jets
\cite{jets, Caporale:2011cc} and Drell-Yan pairs
\cite{Hautmann:2012sh, HHJ1, HHJ2, HHJ3}, and processes with hard
events in both forward and backward direction, {\it i.e.}
forward-backward (`Mueller-Navelet') jets \cite{Colferai:2010wu} and
forward $Z$ boson production combined with a backward jet \cite{Zjet}.

An observable of particular interest is given by forward/backward jets
with a rapidity gap (`Mueller-Tang' jets). This process is special since
it allows to probe the non-forward BFKL kernel, unlike the previously
mentioned processes restricted to the forward case.  From a
phenomenological point of view the description in terms of the  non-forward BFKL Green's function  is of  relevance,  since the latter
 describes a color singlet  $t$-channel exchange. 
 Unlike configurations which merely
suppress emissions above a certain veto scale, the non-forward BFKL Green's function therefore describes a $t$-channel exchange 
with emissions into the gap region  intrinsically absent.

While the non-forward BFKL kernel is currently available at
next-to-leading order (NLO)~\cite{Fadin:2005zj}, only the virtual NLO
corrections to the impact factors \cite{papa} are currently known;
phenomenological studies, see {\it e.g.  } \cite{Kepka:2010hu,
  Enberg:2001ev}, are therefore limited to leading order (LO) impact
factors.  As NLO corrections to BFKL observables are often found to be
size-able, this limitation to LO impact factors is currently one of the
main drawbacks of BFKL phenomenology.  A powerful tool to overcome
this limitation is given by Lipatov's effective
action~\cite{Lipatov:1995pn}.  It is given in terms of the
conventional QCD action to which a new induced term is added. The
latter contains a new effective degree of freedom, the reggeized
gluon, which has been introduced in order to achieve a gauge invariant
factorization of QCD amplitudes in the high energy limit.  The
determination of higher order corrections within this effective action
is at first plagued by both technical and conceptual
difficulties. Loop corrections show a new type of divergence, which is
not present in usual QCD Feynman diagrams. Supplementing the QCD
action with the additional induced term leads to an apparent
over-counting problem. These problems have been addressed and resolved
recently, first in the context of LO transition
kernels~\cite{Hentschinski:2009ga, Hentschinski:2009zz,
  Hentschinski:2009zz,Hentschinski:2008rw,Hentschinski:2008im} and
later on in the calculation of NLO corrections to the forward
quark-initiated jet vertex \cite{quarkjet} and the quark contribution
to the two-loop gluon trajectory \cite{traj}, for a recent review see
\cite{review}.

In this contribution we present some details of the determination of
the missing real NLO correction to the quark-initiated Mueller-Tang
jet impact factors. For further details and the complete result,
including the gluon-initiated jet, we refer
to~\cite{Hentschinski:prep2}.

\section{The high energy effective action}
\label{sec:LODYimpact}
 
   The
effective action adds to the QCD action an induced term, $
S_{\text{eff}} = S_{\text{QCD}} + S_{\text{ind.}}$, which describes
the coupling of the reggeized gluon field $A_\pm(x) = -i t^a A_\pm^a(x)
$ to the usual gluonic field $v_\mu(x) = -it^a v_\mu^a(x)$. This induced term
reads
\begin{align}
\label{eq:1efflagrangian}
  S_{\text{ind.}}[v_\mu, A_\pm]& = \int \! \text{d}^4 x \,
\tr\bigg[\bigg( W_+[v(x)] - A_+(x) \bigg)\partial^2_\perp A_-(x)\bigg]
\notag \\
&  \qquad  \qquad   \qquad
+\int \! \text{d}^4 x \, \tr\bigg[\bigg(W_-[v(x)] - A_-(x) \bigg)\partial^2_\perp A_+(x)\bigg]
.
\end{align}
The infinite number of couplings of the gluon field to the reggeized
gluon field are encoded in two functionals $W_\pm[v]  =
v_\pm \frac{1}{ D_\pm}\partial_\pm $ where $ D_\pm = \partial_\pm + g v_\pm$.
Reggeized gluon fields are invariant under local  SU$(N_c)$ gauge transformations. Strong ordering of longitudinal momenta in high
energy factorized amplitudes provides the following   kinematic
constraint,
\begin{align}
  \label{eq:constraint}
\partial_+ A_-(x)  =  \partial_- A_+(x) = 0,
\end{align} 
which is always implied. 
Quantization of the gluonic field requires to add  gauge fixing and ghost terms, which we have included in  the QCD action.
\begin{figure}[htb]
  \centering
   \parbox{.7cm}{\includegraphics[height = 1.8cm]{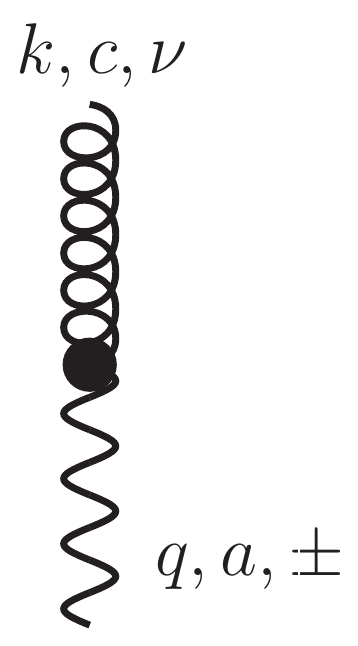}} $=  \displaystyle   \begin{array}[h]{ll}
    \\  \\ - i{\bm q}^2 \delta^{a c} (n^\pm)^\nu,  \\ \\  \qquad   k^\pm = 0.
   \end{array}  $ 
 \parbox{1.2cm}{ \includegraphics[height = 1.8cm]{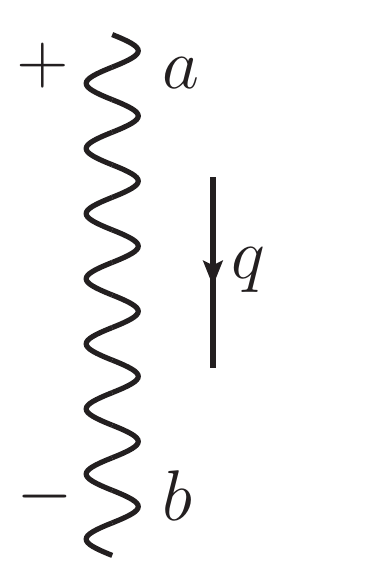}}  $=  \displaystyle    \begin{array}[h]{ll}
    \delta^{ab} \frac{ i/2}{{\bm q}^2} \end{array}$ 
 \parbox{1.7cm}{\includegraphics[height = 1.8cm]{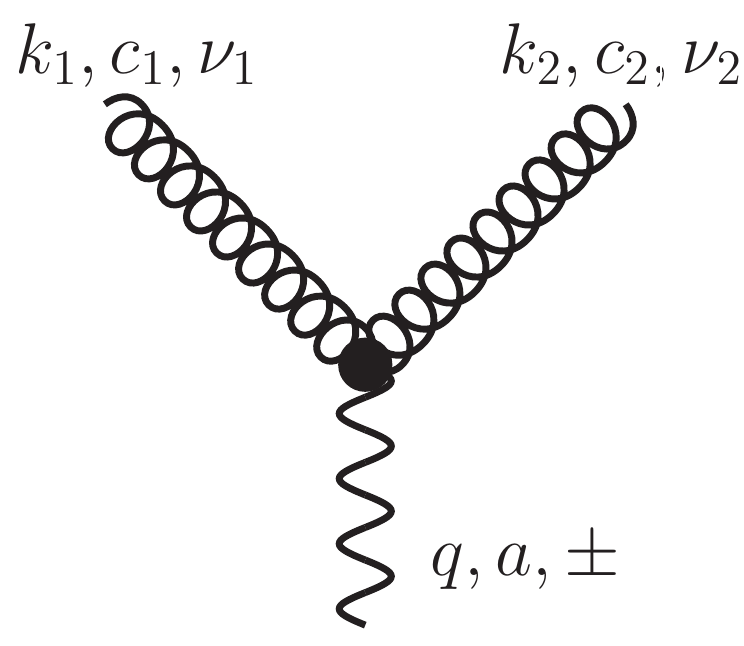}} $ \displaystyle  =  \begin{array}[h]{ll}  \\ \\ g f^{c_1 c_2 a} \frac{{\bm q}^2}{k_1^\pm}   (n^\pm)^{\nu_1} (n^\pm)^{\nu_2},  \\ \\ \quad  k_1^\pm  + k_2^\pm  = 0
 \end{array}$

  \caption{\small From left to the right: the direct transition vertex, the reggeized gluon propagator and the  order $g$ induced vertex.}
  \label{fig:feynrules0p2}
\end{figure}
Feynman rules are given in Fig.~\ref{fig:feynrules0p2}. Curly
lines describe the conventional QCD gluon field and wavy 
lines the reggeized gluon field.  There exist an infinite number of
higher order induced vertices. For the present analysis only the order
$g$ induced vertex in Fig.~\ref{fig:feynrules0p2} is needed. Loop
corrections furthermore require a regularization of the light-cone
singularity $1/k_1^\pm$.  As discussed in~\cite{Hentschinski:2011xg}
this pole should be treated as a Cauchy principal value.

\section{Mueller-Tang impact factors}
\begin{figure}[htb]
  \centering
  \parbox{3.7cm}{\includegraphics[width = 3.7cm]{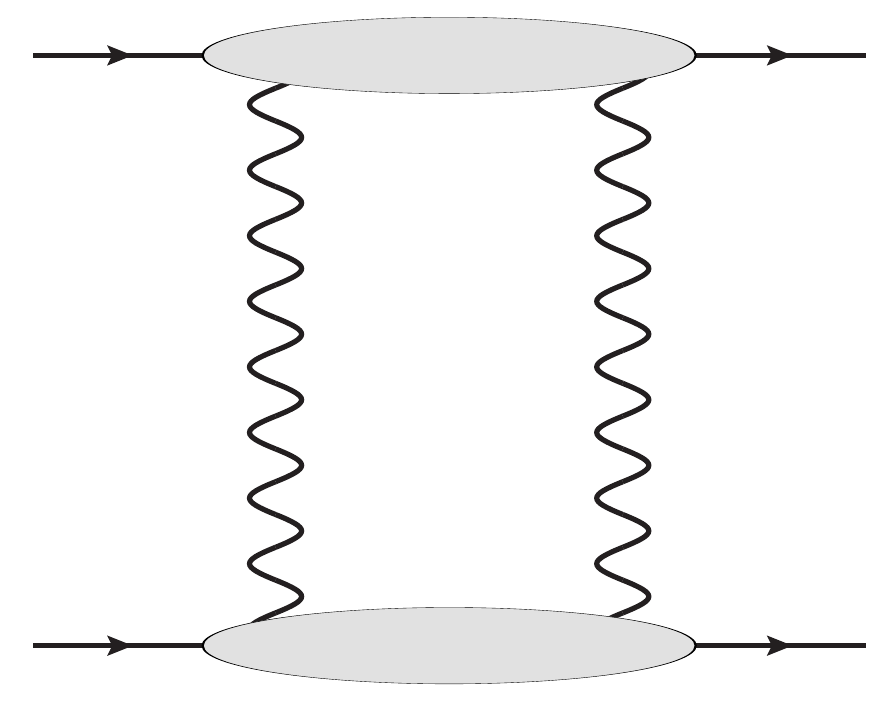}} 
$\displaystyle
\begin{array}[h]{l}
 \parbox{3cm}{\includegraphics[width = 3cm]{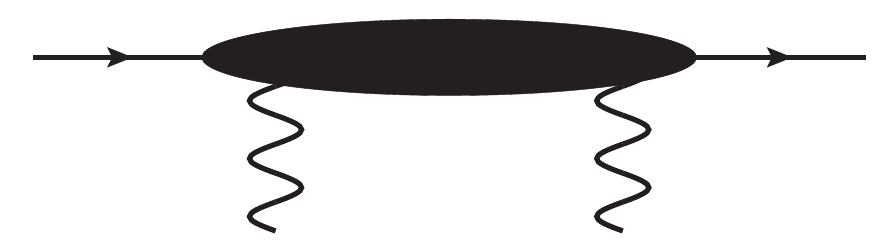}}  \qquad  \parbox{3cm}{\includegraphics[width = 3cm]{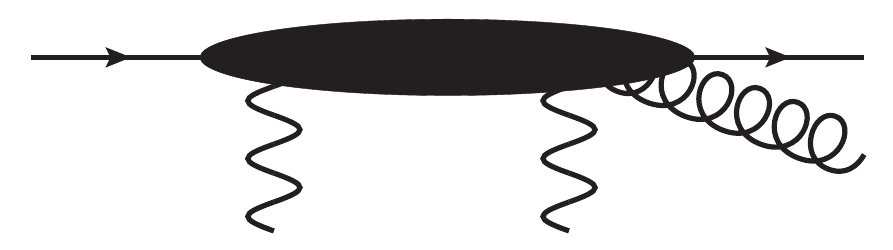}}   \\ \\ \\ 

 \parbox{3cm}{\includegraphics[width = 3cm]{mt_lo.pdf}}   = 
\parbox{3cm}{\includegraphics[width = 3cm]{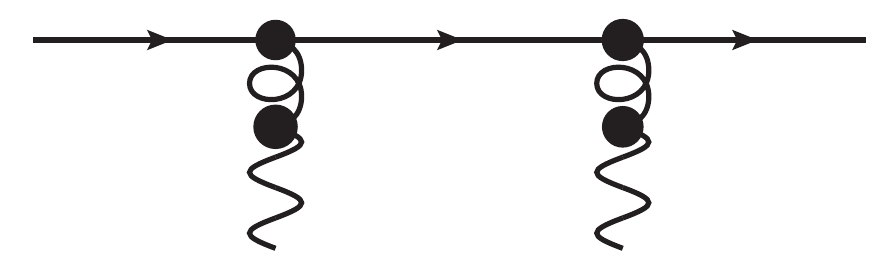}}   
+
\parbox{3cm}{\includegraphics[width = 3cm]{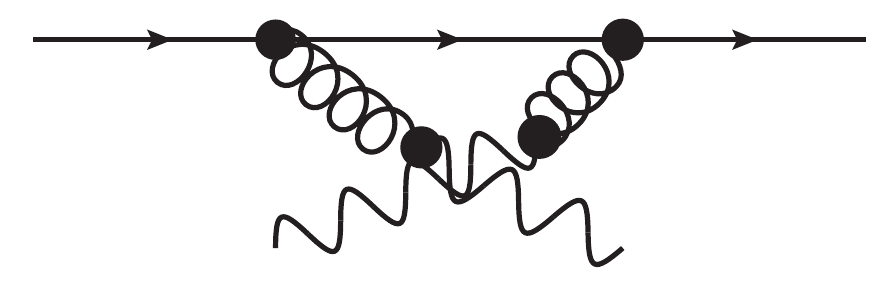}} 
  
\end{array}
$
  \caption{\small Left: Quark-quark scattering amplitude with exchange of two reggeized gluons in the color singlet. Right, top: leading and next-to-leading (real corrections) Mueller-Tang impact factor. Right, bottom: diagrammatic expansion of the LO impact factor.}
  \label{fig:mt}
\end{figure}

The starting point for the determination of the quark induced Mueller-Tang
jet impact factors is given by the quark-quark scattering amplitude
with color singlet exchange. In the high energy limit such a
scattering amplitude factorizes (in terms  of a transverse convolution
integral) into the two reggeized gluon exchange in the color singlet
(which can be understood as the lowest order contribution to the
perturbative Pomeron) and two impact factors (which describe the
coupling of the two reggeized gluon state to the external quarks), see
Fig.~\ref{fig:mt}.
Due to high energy factorization, the longitudinal part of the loop integral {\it i.e.} the  $l^-$ ($l^+$) integration,   can be entirely  associated  with the upper (lower) impact factor. At cross-section level, using dimensional regularization in $d = 4 + 2 \epsilon$ dimensions,  the  leading order result reads, 
\begin{align}
  \label{eq:Born}
  d \sigma_{ab} &=  H^{(0)}_a H^{(0)}_b \left[ \int \frac{d^{2 + 2\epsilon} {\bm l}_1}{\pi^{1 + \epsilon}} \frac{1}{{\bm l}_1^2 ({\bm k} - {\bm l}_1)} \right]\left[ \int \frac{d^{2 + 2\epsilon} {\bm l}_2}{\pi^{1 + \epsilon}} \frac{1}{{\bm l}_2^2 ({\bm k} - {\bm l}_2)} \right] d^{2 + 2\epsilon} {\bm k}
\end{align}
with the impact factor
\begin{align}
  \label{eq:H0}
  H^{(0)}_q & = \frac{C_f^2}{N_c^2 -1}  \frac{N_c}{2 N_c}  \tr  (\fdag{p} \fdag{n} \fdag{p_a} \fdag{n} )  \frac{1}{2 p_a^+}   \cdot   \frac{1}{p_a^+}\int \frac{d k^-}{2 \pi} 2\pi \delta (k^- - \frac{{\bm p}^2}{p_a^+}) \frac{g^4 }{2(4 \pi)^{2 +2 \epsilon}} 
\notag \\
&=
\frac{\alpha_s^2 C_f^2  }{\mu^{4\epsilon}\Gamma^2(1-\epsilon)(N_c^2 -1)},
\end{align}
in agreement with \cite{Mueller:1992pe}. While the leading order
impact factor is merely a constant, the real next-to-leading order
corrections depend both on the momenta of the final state particles and the loop momentum of the reggeized gluon loop, ${\bm l}_1$ and ${\bm l}_2$.  Their integrated version  reads
\begin{align}
  \label{eq:resultss}
& H^{(1)}_{r} =
\int_0^1 d z \int \frac{d^{2 + 2 \epsilon} {\bm q}}{\pi^{1 + \epsilon}} 
H^{(0)}
\frac{\alpha_s}{ 2\pi }
 \frac{ P_{qg}(z, \epsilon)}{\Gamma(1-\epsilon) \mu^{2\epsilon}}
 \bigg[   C_f   
 \left(
\frac{{\bm \Delta}}{{\bm \Delta}^2} - \frac{{\bm q}}{{\bm q}^2}
\right)
\notag 
\\
& -
{C_a}\left( \frac{{\bm p}}{{\bm p}^2} + \frac{1}{2} \frac{{\bm \Sigma}_1}{{\bm \Sigma}_1^2} + \frac{1}{2} \frac{{\bm \Upsilon}_1}{{\bm \Upsilon}^2_1} \right)
  \bigg]
 \left[  C_f    
\left(
\frac{{\bm \Delta}}{{\bm \Delta}^2} - \frac{{\bm q}}{{\bm q}^2}
\right)
-
{C_a}{}\left( \frac{{\bm p}}{{\bm p}^2} + \frac{1}{2} \frac{{\bm \Sigma}_2}{{\bm \Sigma}_2^2} + \frac{1}{2} \frac{{\bm \Upsilon}_2}{{\bm \Upsilon}^2_2} \right)
  \right]
\end{align}
where  $ {\bm \Sigma}_i  = {\bm q} - {\bm l}_i$, $ {\bm \Upsilon}_i  = {\bm q} - {\bm k} + {\bm l}_i$, and $i= 1,2$. ${\bm k}$  is the momentum transfer in the $t$-channel,  ${\bm q}$ the transverse momentum of the final state gluon. $   P_{qg}(z,\epsilon)$ is 
 the real part of the $q \to g$ splitting function. For the limit $z \to 0$, the above expression can be shown to agree with the real part of the triple-Pomeron vertex \cite{Bartels:1994jj, Hentschinski:2009zz}. 


\begin{theacknowledgments}
We would like to thank   G.~Chachamis, J.~Madrigal~Mart\'inez and A.~Sabio~Vera for fruitful collaboration.   M.H. acknowledges support from  the U.S. Department of Energy under contract number DE-AC02-98CH10886 and a BNL ``Laboratory Directed Research and Development'' grant (LDRD 12-034).
\end{theacknowledgments}



\bibliographystyle{aipproc}   


\IfFileExists{\jobname.bbl}{}
 {\typeout{}
  \typeout{******************************************}
  \typeout{** Please run "bibtex \jobname" to optain}
  \typeout{** the bibliography and then re-run LaTeX}
  \typeout{** twice to fix the references!}
  \typeout{******************************************}
  \typeout{}
 }

\end{document}